\documentclass{article}
\usepackage{latexsym}
\usepackage{graphicx}

\input{psfig.sty}

\begin{document}

\begin{flushright}
\large NUMI-L-713 \\
\large hep-ph/0101090
\end{flushright}

\vspace{10mm}

\begin{center}
\LARGE Matter enhanced $\nu_\mu\rightarrow\nu_e$ signals using
various Fermilab main injector beam configurations
\end{center}

\begin{center}
\large Lawrence Wai and Bradley Patterson \\
Stanford University 
\end{center}

\begin{center}
\large January 17, 2001
\end{center}

\vspace{10mm}

\section{Introduction}

\begin{figure}
\centerline{\psfig{file=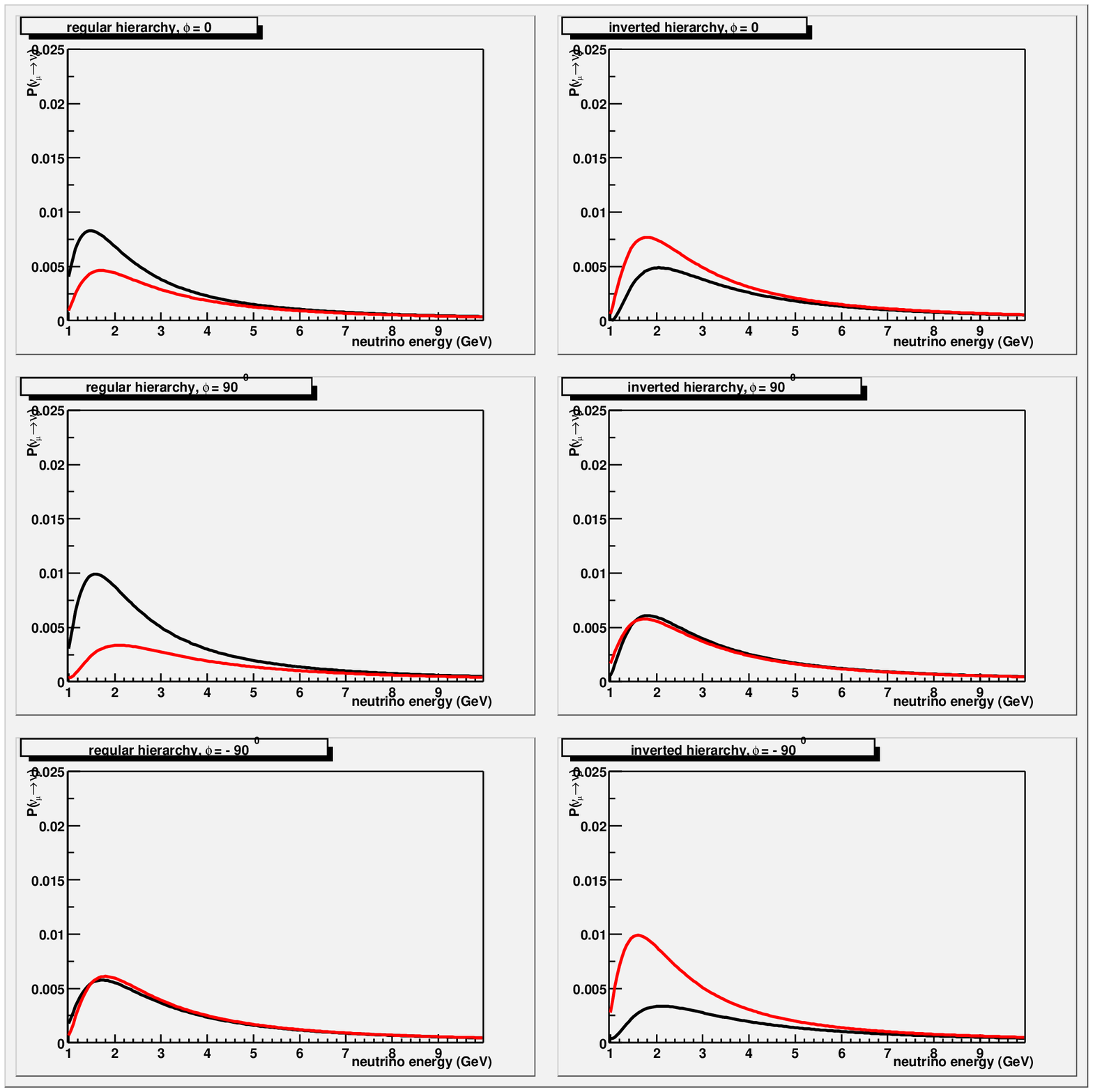}}
\caption{Probability for $\nu_\mu\rightarrow\nu_e$ for FNAL-Soudan
(L=732km), given $|U_{e3}|^2=0.003$ (ie. $\theta_{13}=3^o$), $\Delta
m_{23}^2=0.003eV^2$, $\Delta m_{12}^2=0.00003eV^2$, $U_{\mu
3}^2=U_{\tau 3}^2$, and $U_{e 1}^2=U_{e 2}^2$.  The black lines are
for neutrinos, and the red lines are for anti-neutrinos.  The panels
on the left side have regular mass hierarchy, and the panels on the
right side have inverted mass hierarchy.  The upper panels have zero
phase, the middle panels have +90 degrees phase, and the lower panels
have -90 degrees phase.}
\label{fig_a01}
\end{figure}

\begin{figure}
\centerline{\psfig{file=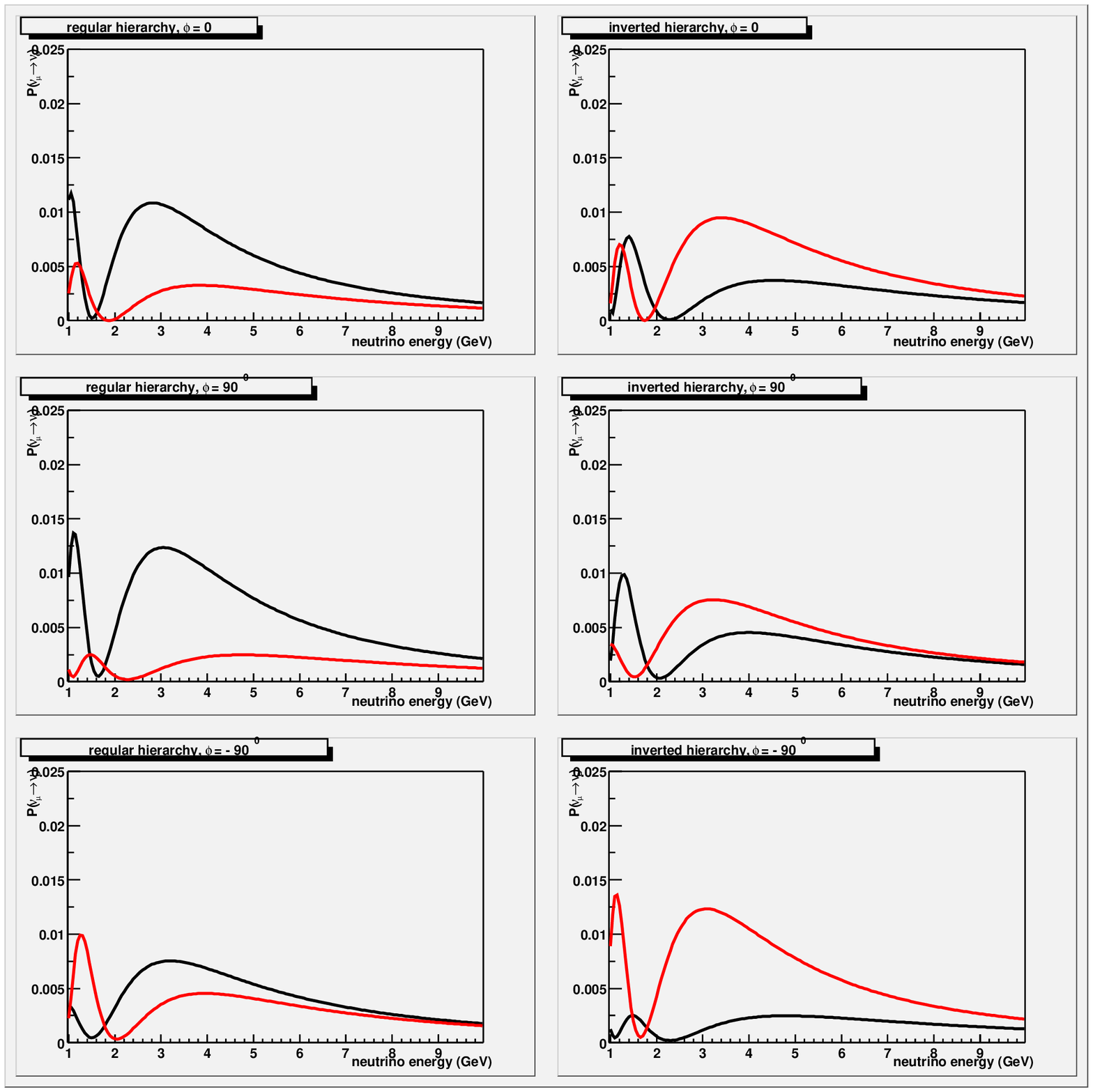}}
\caption{Probability for $\nu_\mu\rightarrow\nu_e$ for FNAL-BNL
(L=1500km), given $|U_{e3}|^2=0.003$ (ie. $\theta_{13}=3^o$), $\Delta
m_{23}^2=0.003eV^2$, $\Delta m_{12}^2=0.00003eV^2$, $U_{\mu
3}^2=U_{\tau 3}^2$, and $U_{e 1}^2=U_{e 2}^2$.  The black lines are
for neutrinos, and the red lines are for anti-neutrinos.  The panels
on the left side have regular mass hierarchy, and the panels on the
right side have inverted mass hierarchy.  The upper panels have zero
phase, the middle panels have +90 degrees phase, and the lower panels
have -90 degrees phase.}
\label{fig_a02}
\end{figure}

\begin{figure}
\centerline{\psfig{file=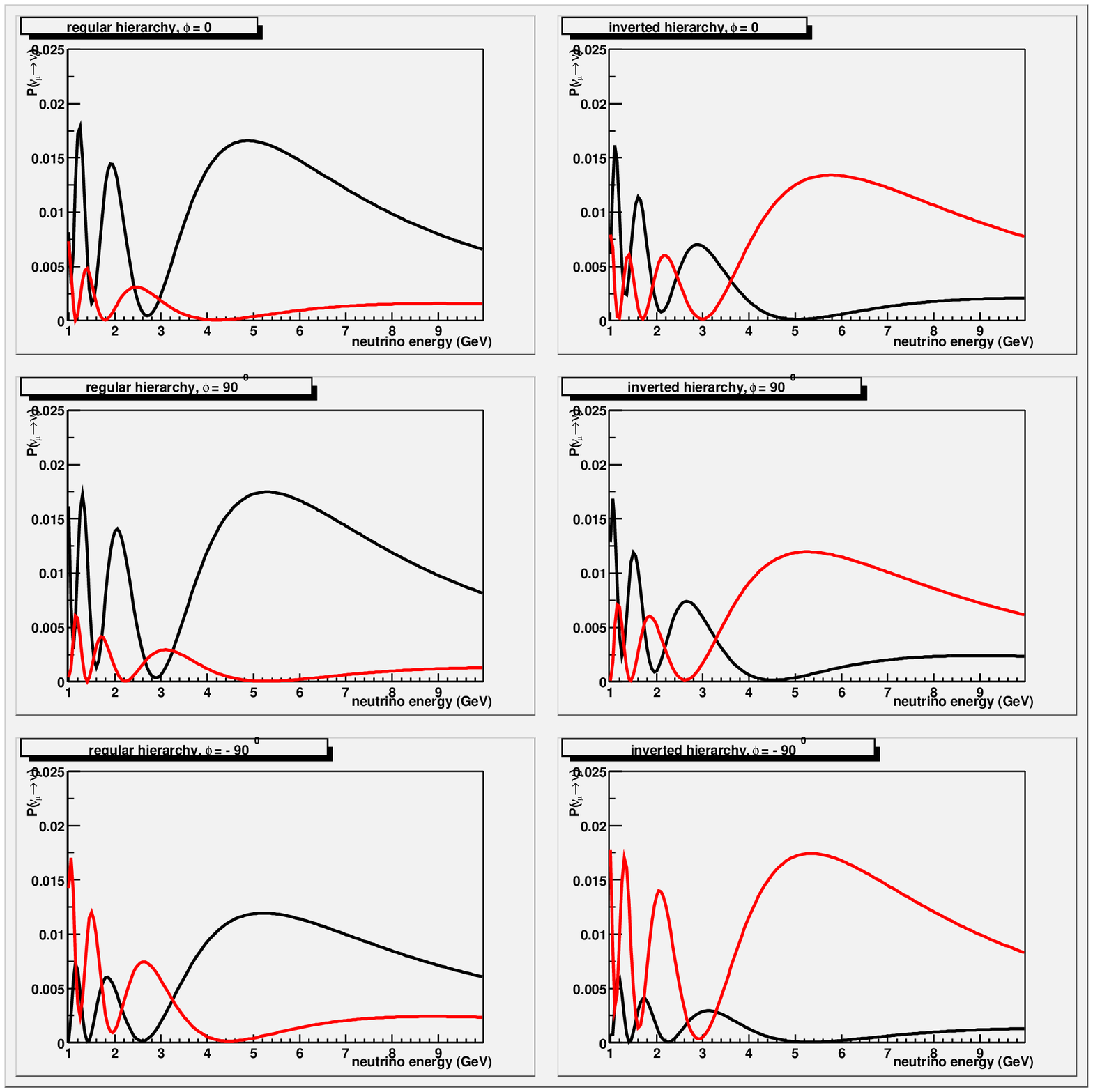}}
\caption{Probability for $\nu_\mu\rightarrow\nu_e$ for FNAL-SLAC
(L=2900km), given $|U_{e3}|^2=0.003$ (ie. $\theta_{13}=3^o$), $\Delta
m_{23}^2=0.003eV^2$, $\Delta m_{12}^2=0.00003eV^2$, $U_{\mu
3}^2=U_{\tau 3}^2$, and $U_{e 1}^2=U_{e 2}^2$.  The black lines are
for neutrinos, and the red lines are for anti-neutrinos.  The panels
on the left side have regular mass hierarchy, and the panels on the
right side have inverted mass hierarchy.  The upper panels have zero
phase, the middle panels have +90 degrees phase, and the lower panels
have -90 degrees phase.}
\label{fig_a03}
\end{figure}

In this report we estimate the sensitivity of experiments optimized
for measuring $|U_{e3}|^2$ and the sign of $\Delta m_{23}^2$ in
various Fermilab main injector beam configurations.  We calculate the
sensitivity to $\nu_\mu\rightarrow\nu_e$ appearance making the
assumption of 3 neutrino generations, and that only solar and
atmospheric neutrino oscillations signals are "real."  We have
implemented an exact calculation of oscillation probabilities in
matter from Barger et.al.~\cite{bar80} for the FNAL-Soudan (732km),
and hypothetical FNAL-BNL (1500km) and FNAL-SLAC (2900km) beamlines,
as shown in figures~\ref{fig_a01},~\ref{fig_a02}, and~\ref{fig_a03}.
Of course, these appearance probabilities must be folded into the
neutrino flux spectrum produced by a particular Fermilab main injector
neutrino beam configuration, as well as the cross-section in the far
detector.  For the neutrino flux spectrum we have used the GEANT
simulation of the NuMI beamline~\cite{hylen97}, and for the neutrino
cross-section we have used the Soudan 2~\cite{soudan00} event
generator.  For the purposes of illustration, we shall use the test
point $\Delta m_{23}^2=0.003eV^2$ (regular hierarchy),
$|U_{e3}|^2=0.003$ (ie. $\theta_{13}=3^o$), $U_{\mu 3}^2=U_{\tau
3}^2$, $\Delta m_{solar}^2=0.00003eV^2$, $U_{e1}^2=U_{e2}^2$, and
phase $\phi=0$.  Using these parameters, the numbers of
$\nu_\mu\rightarrow\nu_e$ charged current interactions per
kiloton-year (in steel) are as follows:
\begin{center}
\begin{tabular}{|l||r|r|r|}
\hline 
\multicolumn{4}{|c|}{{\bf Event rates for steel exposure}} \\
\multicolumn{4}{|c|}{regular hierarchy, $\Delta m_{23}^2=0.003eV^2$, $|U_{e3}|^2=0.003$} \\
\hline 
& FNAL-Soudan & FNAL-BNL & FNAL-SLAC \\ 
\hline 
\hline 
PH2 low $\nu_\mu\rightarrow\nu_e$ CC & 0.93 / kt-yr & 0.55 / kt-yr & 0.23 / kt-yr \\ 
\hline 
PH2 medium $\nu_\mu\rightarrow\nu_e$ CC & 1.41 / kt-yr & 1.15 / kt-yr & 0.79 / kt-yr \\
\hline 
PH2 high $\nu_\mu\rightarrow\nu_e$ CC & 0.98 / kt-yr & 0.95 / kt-yr & 0.87 / kt-yr \\ 
\hline
\end{tabular}
\end{center}

Note that the peak energy values for the PH2 low, medium, and high
energy beams are roughly 3, 6, and 12~GeV respectively.  In this
report we will consider two examples of detectors: a 10-kt
``MINOS-like'' detector, and a 10-kt ``OPERA-like'' detector optimized
for $\nu_e$ appearance.  We define ``MINOS-like detector'' in this
report to be a 10-kt detector composed of alternating 2.54cm Fe plates
and scintillator strips.  We define the ``OPERA-like'' detector as
follows.

In an earlier note, we have described how a hybrid emulsion detector
(HED) can obtain low background measurements of
$\nu_\mu\rightarrow\nu_\tau$ and $\nu_\mu\rightarrow\nu_e$ oscillation
events in the NuMI beam~\cite{wai99b}.  We now consider a HED designed
exclusively to measure $\nu_e$.  The tracking requirements for
$\nu_\tau$ appearance are more stringent than those for $\nu_e$
appearance, and thus we can significantly relax certain parameters of
a HED optimized for $\nu_\tau$~\cite{wai99a} if we only care about
identifying $\nu_e$.  The main difference is that we propose to use
5~mm~Fe instead of 1~mm~Pb as in OPERA~\cite{opera00}.  Thus, for
roughly the same emulsion cost as in OPERA (a 2-kt detector) we obtain
a detector with 3.5 times greater mass.

The detector would be composed of 10~cm thick steel-emulsion stacks,
separated by RPC planes.  The RPC planes would be used for triggering
and event location.  Emulsion sheets would be sandwiched between 5~mm
thick steel plates.  If the planes have an area of 8x8~square~meters,
then the mass of one steel-emulsion target plane would be 50~tons.
The entire detector would consist of 200 planes.  Note that there is
enough floor space behind the first two MINOS supermodules in the
MINOS Soudan cavern for such a detector.

We now attempt to estimate how much a HED would cost.  We use the cost
of emulsion film from the OPERA proposal~\cite{opera00} (converted to
US dollars).  We estimate the cost of salary, wages, and institutional
overhead to be the same fraction as in the MINOS TDR~\cite{tdr98}, ie
50\% of total cost before contingency.  We add a 50\% contingency.
This results in a total detector cost of roughly \$18.4M per kt.  The
rough cost breakdown is shown in the following table:
\begin{center}
\begin{tabular}{|l||r|r|r|}
\hline
\multicolumn{4}{|c|}{Cost breakdown for 10-kt HED} \\
\hline
component & cost per unit & \# units & total cost \\
\hline
\hline
emulsion film & \$150 / $m^2$ & 256,000 & \$38.4M \\
\hline
steel & \$1000 / ton & 10,000 & \$10.0M \\
\hline
RPC+electronics & \$500 / $m^2$ & 25,600 & \$12.8M \\
\hline
wages, salary, overhead & & & \$61.2M \\
\hline
contingency & & & \$61.2M \\
\hline
\hline
total & & & \$183.6M \\
\hline
\end{tabular}
\label{cost}
\end{center}

We now turn to a comparison of MINOS-like and OPERA-like detector
analyses.

\section{MINOS-like data analysis}

\begin{figure}
\centerline{\psfig{file=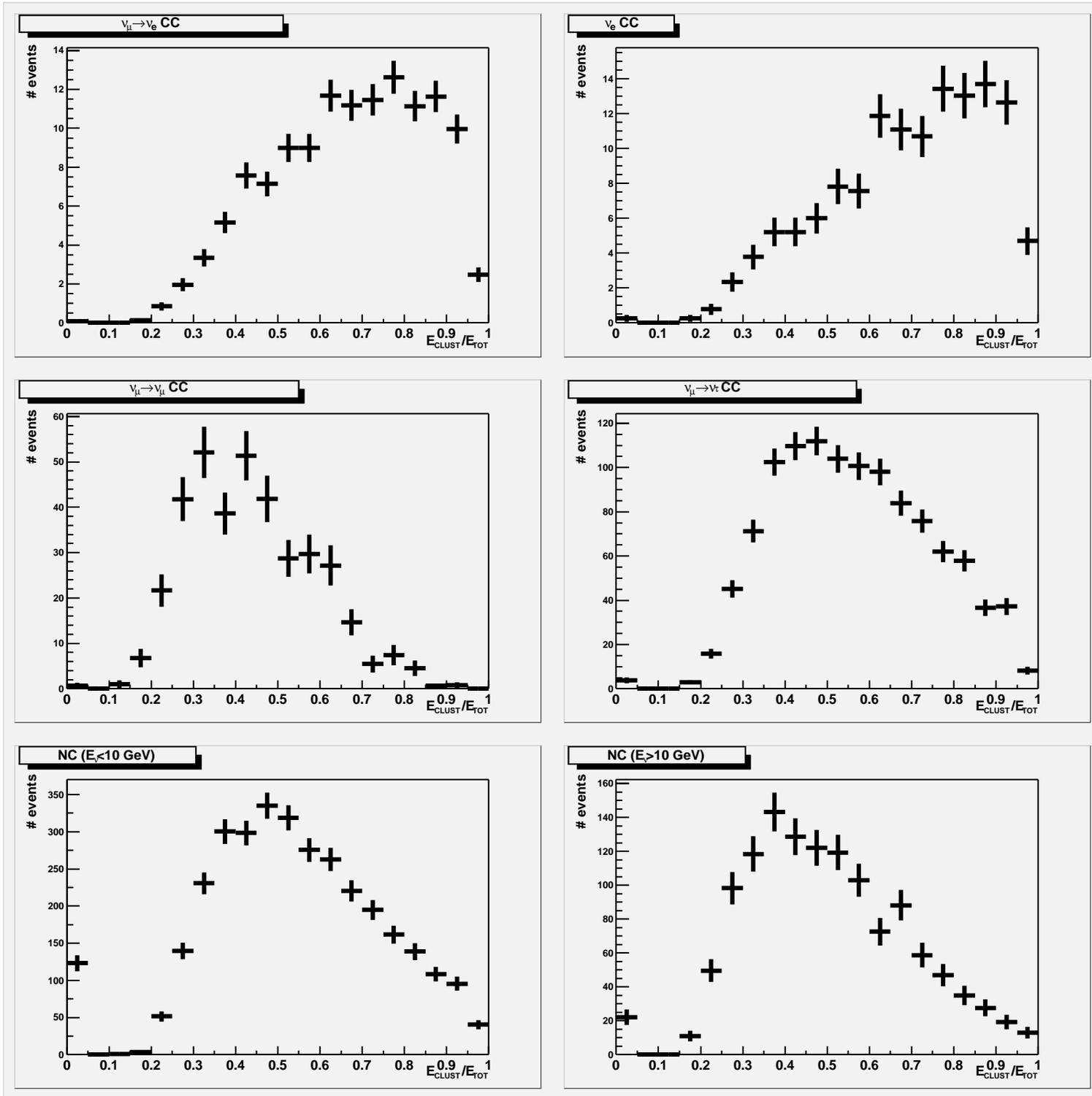}}
\caption{PH2 medium/MINOS-like/SLAC event distributions of the
fraction of total energy in the highest energy cluster
($E_{CLUST}/E_{TOT}$).  We require $E_{CLUST}/E_{TOT}>0.7$.  The
histograms are normalized to 4x 40 kt-yr data samples.  An exact
calculation of the oscillation probability in matter was used for
$\Delta m_{23}^2=0.003eV^2$ (regular mass hierarchy),
$|U_{e3}|^2=0.003$ (ie. $\theta_{13}=3^o$), $U_{\mu 3}^2=U_{\tau
3}^2$, $U_{e 1}^2=U_{e 2}^2$, $\Delta m_{12}^2=0.00003eV^2$, and phase
$\phi=0$.}
\label{fig_d01}
\end{figure}

\begin{figure}
\centerline{\psfig{file=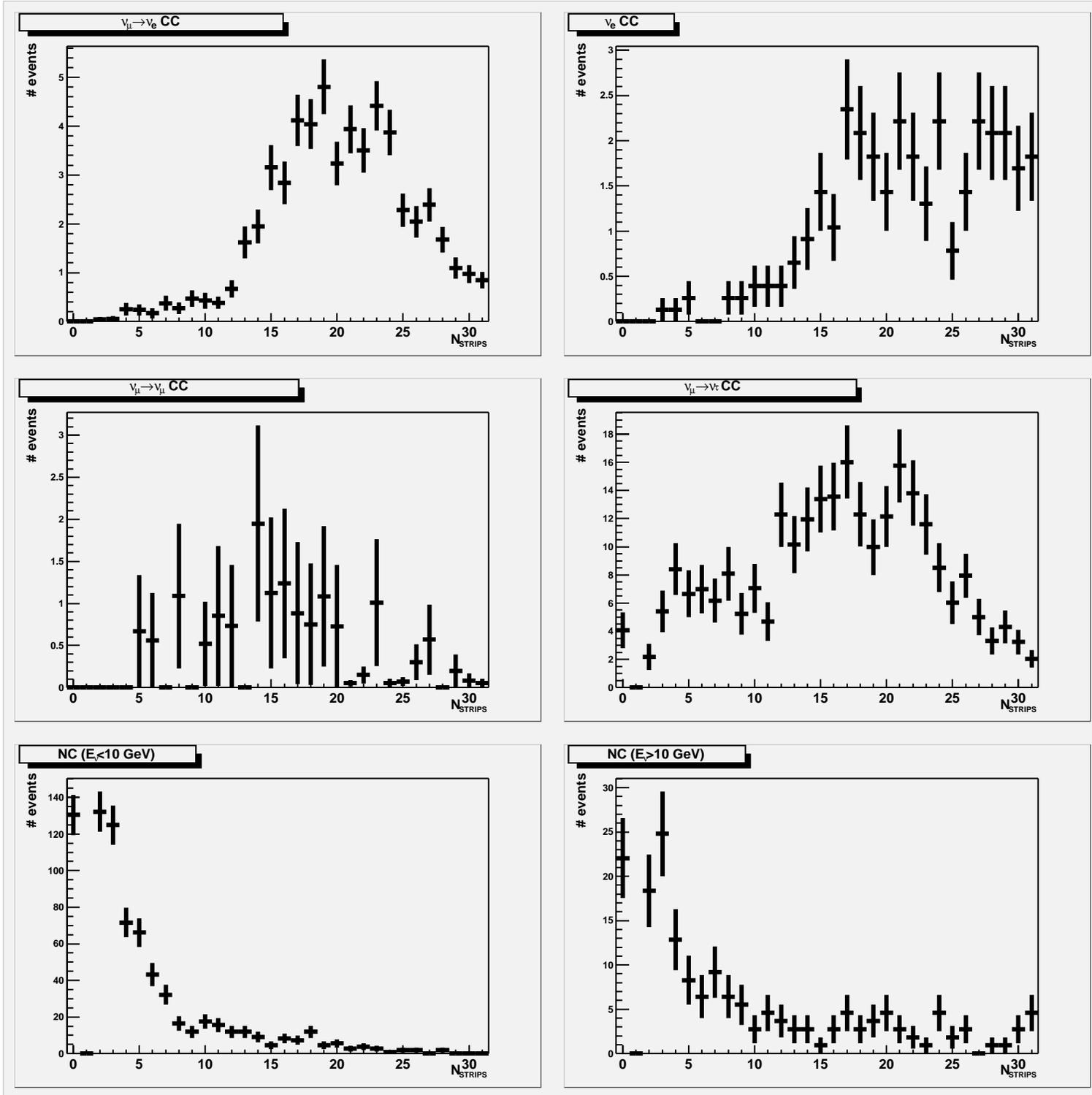}}
\caption{PH2 medium/MINOS-like/SLAC event distributions for number of
strips ($N_{STRIPS}$) in the highest energy cluster.  We require
$N_{STRIPS}>=9$.  The histograms are normalized to 4x 40 kt-yr data
samples.  An exact calculation of the oscillation probability in
matter was used for $\Delta m_{23}^2=0.003eV^2$ (regular mass
hierarchy), $|U_{e3}|^2=0.003$ (ie. $\theta_{13}=3^o$), $U_{\mu
3}^2=U_{\tau 3}^2$, $U_{e 1}^2=U_{e 2}^2$, $\Delta
m_{12}^2=0.00003eV^2$, and phase $\phi=0$.}
\label{fig_d02}
\end{figure}

\begin{figure}
\centerline{\psfig{file=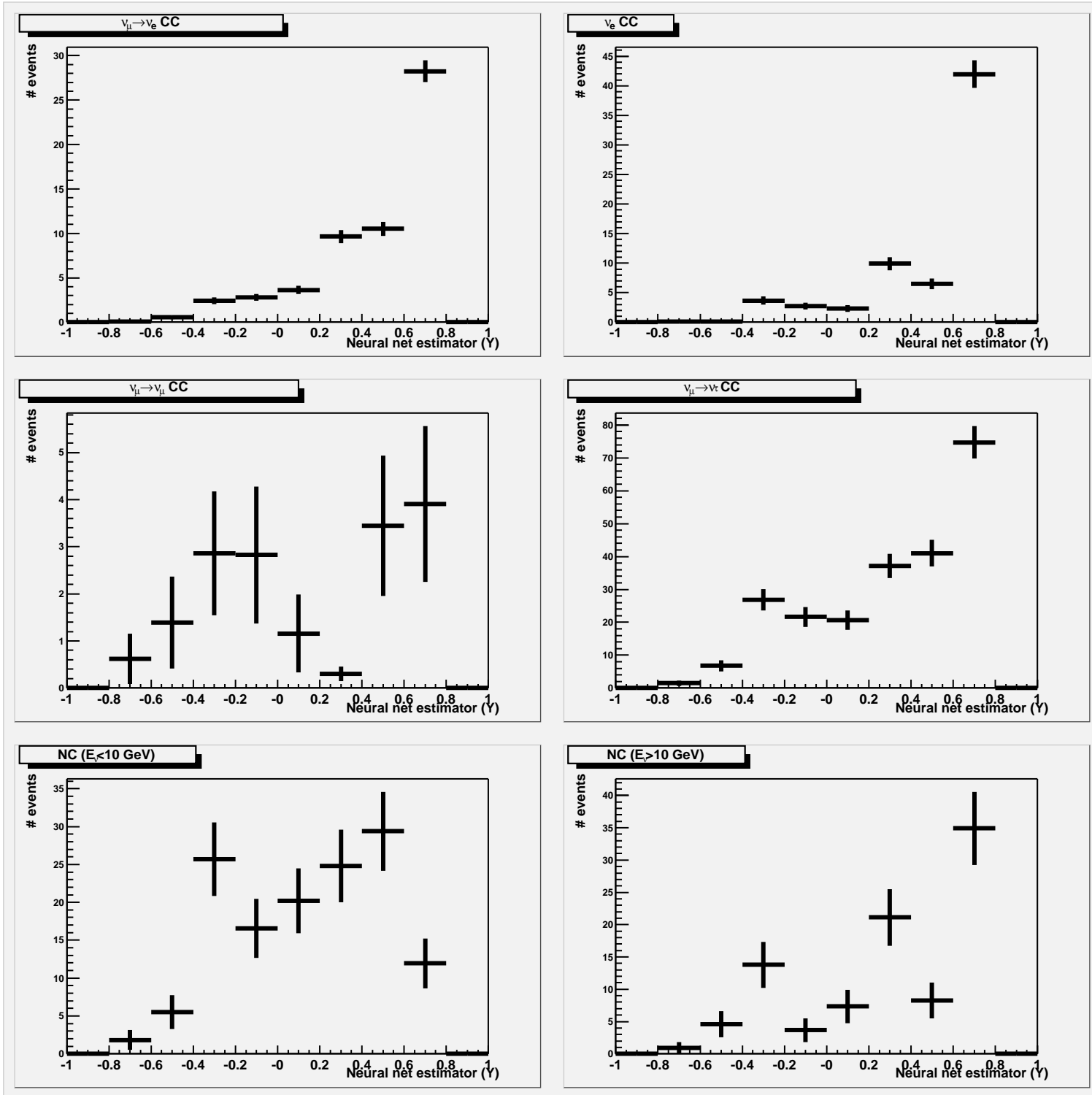}}
\caption{PH2 medium/MINOS-like/SLAC event distributions for the neural
net estimator ($Y$). We require $Y>0$.  The histograms are normalized
to 4x 40 kt-yr data samples.  An exact calculation of the oscillation
probability in matter was used for $\Delta m_{23}^2=0.003eV^2$
(regular mass hierarchy), $|U_{e3}|^2=0.003$ (ie. $\theta_{13}=3^o$),
$U_{\mu 3}^2=U_{\tau 3}^2$, $U_{e 1}^2=U_{e 2}^2$, $\Delta
m_{12}^2=0.00003eV^2$, and phase $\phi=0$.}
\label{fig_d03}
\end{figure}

\begin{figure}
\centerline{\psfig{file=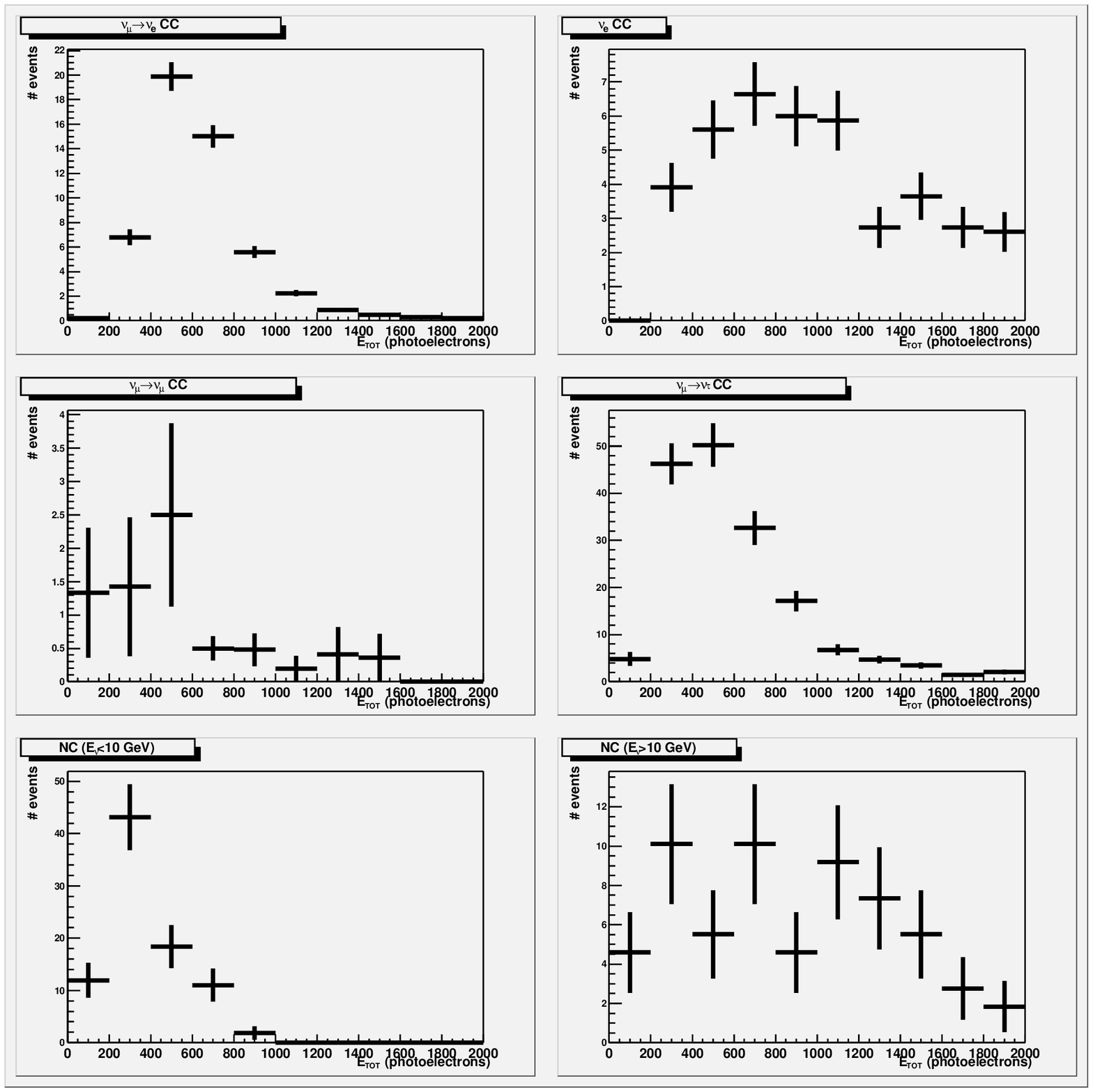}}
\caption{PH2 medium/MINOS-like/SLAC event distributions for the total
energy ($E_{TOT}$).  We require 400 photoelectrons $ < E_{TOT} < $ 800
photoelectons.  The histograms are normalized to 4x 40 kt-yr data
samples.  An exact calculation of the oscillation probability in
matter was used for $\Delta m_{23}^2=0.003eV^2$ (regular mass
hierarchy), $|U_{e3}|^2=0.003$ (ie. $\theta_{13}=3^o$), $U_{\mu
3}^2=U_{\tau 3}^2$, $U_{e 1}^2=U_{e 2}^2$, $\Delta
m_{12}^2=0.00003eV^2$, and phase $\phi=0$.}
\label{fig_d04}
\end{figure}

\begin{figure}
\centerline{\psfig{file=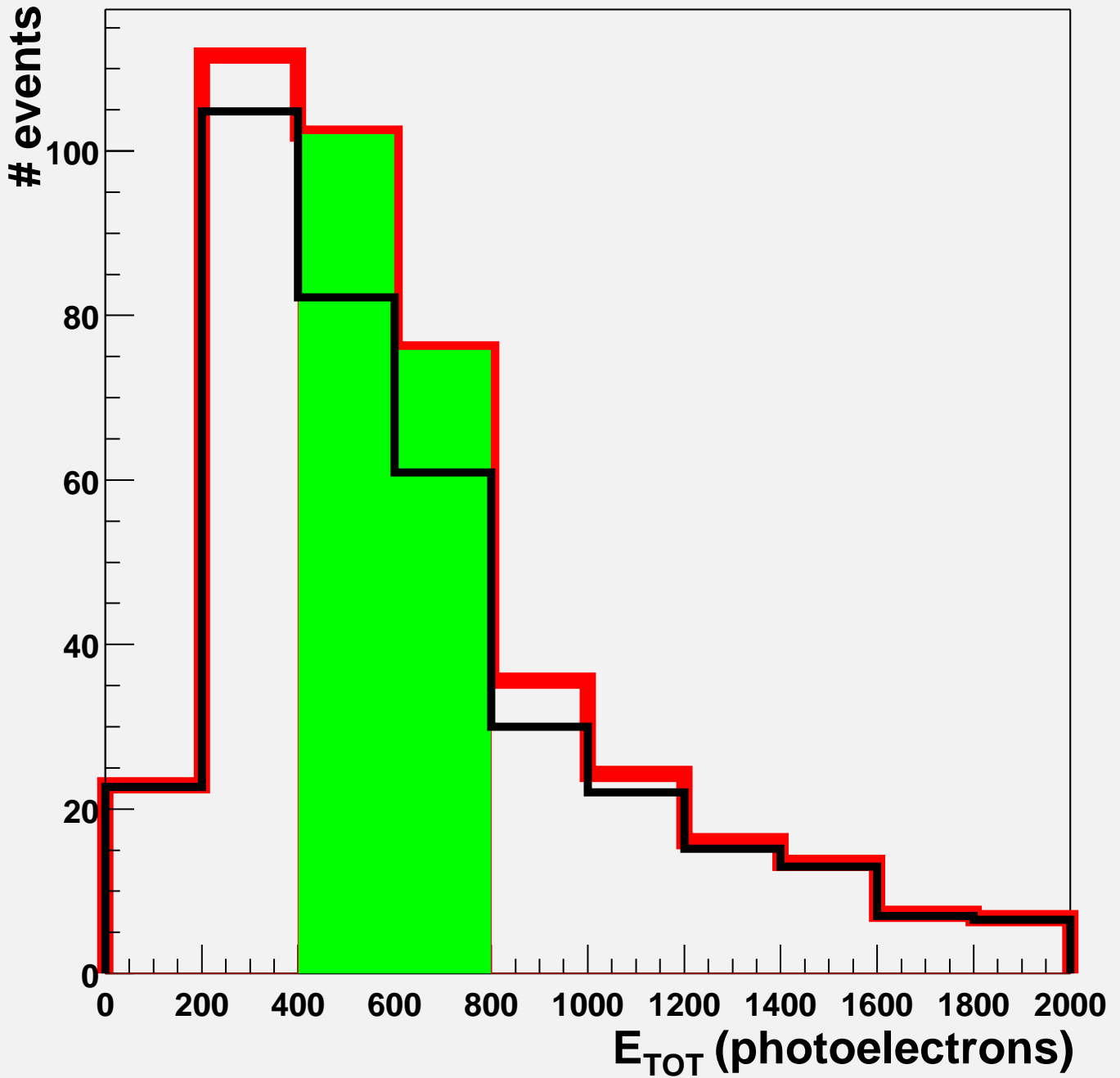}}
\caption{PH2 medium/MINOS-like/SLAC total energy event distributions
for background (black line) and background plus signal (red line).  We
require the total energy to be in the range between 400 and 800
photoelectons (green region).  The histograms are normalized to 4x 40
kt-yr data samples.  An exact calculation of the oscillation
probability in matter was used for $\Delta m_{23}^2=0.003eV^2$
(regular mass hierarchy), $|U_{e3}|^2=0.003$ (ie. $\theta_{13}=3^o$),
$U_{\mu 3}^2=U_{\tau 3}^2$, $U_{e 1}^2=U_{e 2}^2$, $\Delta
m_{12}^2=0.00003eV^2$, and phase $\phi=0$.}
\label{fig_d05}
\end{figure}

The MINOS-like detector analysis (taken from~\cite{wai00}) is based
upon the full ``official'' GEANT simulation of the MINOS
detector~\cite{hatcher00}.  We use the case of PH2 medium to SLAC as
an example.  The basic strategy can be summarized as follows: 1)
reject events with $P_{\mu}>1GeV$, 2) fraction of energy in the
highest energy cluster $E_{CLUST}/E_{TOT}>0.7$ (see
figure~\ref{fig_d01}), 3) number of strips in the highest energy
cluster $N_{STRIPS}>=9$ (see figure~\ref{fig_d02}), 4) neural net
estimator consistent with $\nu_e$ (see figure~\ref{fig_d03}), and 5)
reject events in which the total energy ($E_{TOT}$) does not fall
within some energy range optimized to $\Delta m^2$ (see
figures~\ref{fig_d04} and~\ref{fig_d05}).  The numbers of events after
each sequential cut are tabulated in table~\ref{tab_s01}.

\section{HED data analysis}

The HED data analysis~\cite{wai99b} is based upon simple gaussian
smearing of individual Monte Carlo particle truth information.  Of
course, the smearing is appropriate to the HED detector under
consideration.  The basis for the inputs to the smearing come from the
OPERA proposal~\cite{opera00} (which has an $X_0$=0.18 for each target
plate), with the appropriate extrapolation to the thicker plates in
the HED described in this report ($X_0$=0.28).  The resolution on
electromagnetic showers from track counting in the OPERA proposal is
20\%/sqrt(E), so we use a gaussian smearing of 25\%/sqrt(E) for
electrons and gammas in our calculation.  The resolution on charged
particle momentum using the multiple Coulomb scattering method in the
OPERA proposal is 16\%, so we use a gaussian smearing of 20\% to
(non-electron) charged particles in our calculation.  We use a
1~milliradian angular resolution for charged particle direction
(assuming that we use the emulsion film described in the OPERA
proposal) and we add in quadrature the angular smearing due to
multiple coulomb scattering in the target steel plate (using a
randomly chosen scattering vertex position in the target plate).

\begin{figure}
\centerline{\psfig{file=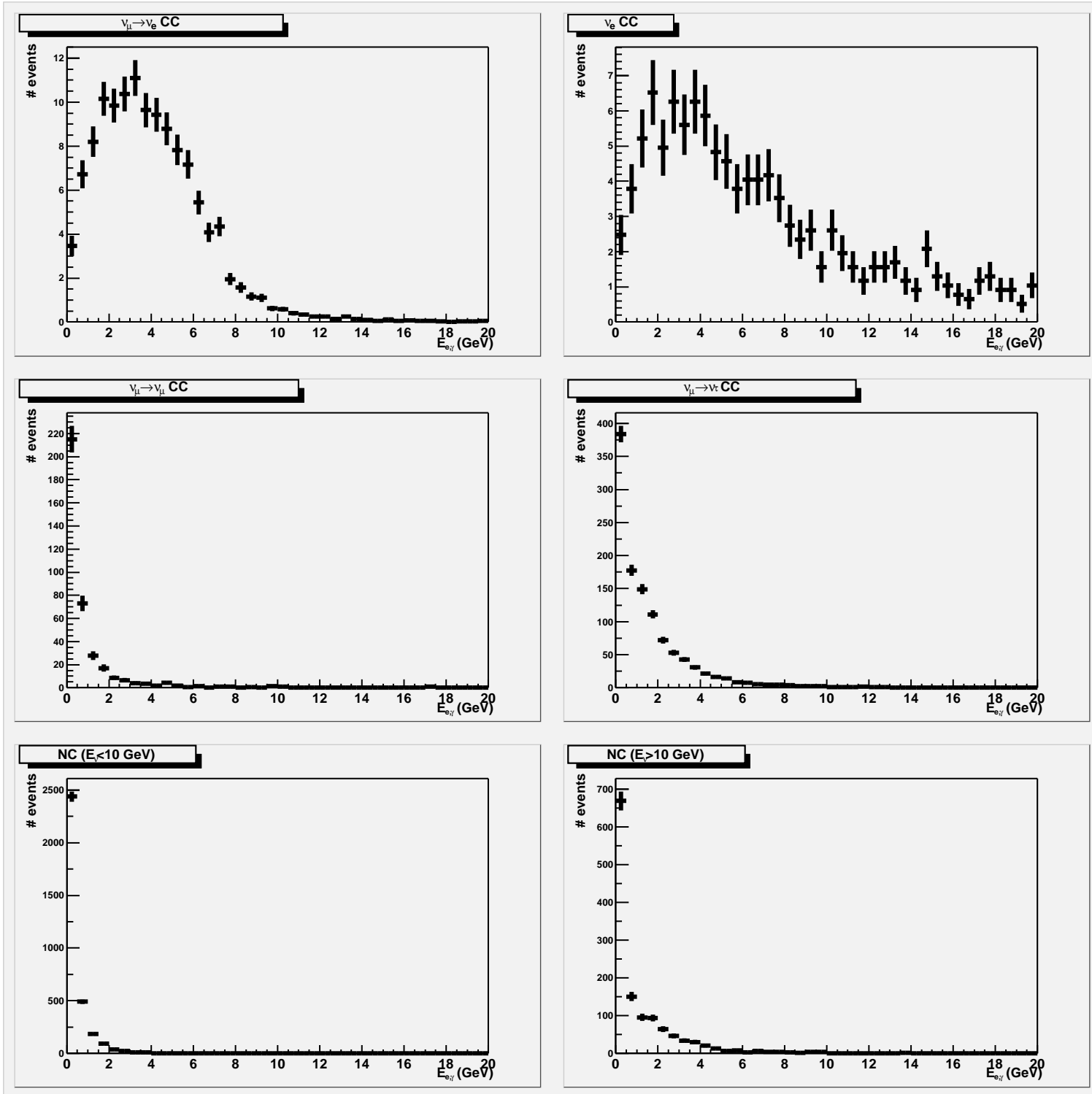}}
\caption{PH2 medium/HED/SLAC event distributions for the maximum
energy electromagnetic shower ($E_{e,\gamma}$).  We have used a
25\%/sqrt(E) electromagnetic shower energy resolution (from track
counting).  We require $E_{e,\gamma} > $~5~GeV.  The histograms are
normalized to 4x 40 kt-yr data samples.  An exact calculation of the
oscillation probability in matter was used for $\Delta
m_{23}^2=0.003eV^2$ (regular mass hierarchy), $|U_{e3}|^2=0.003$
(ie. $\theta_{13}=3^o$), $U_{\mu 3}^2=U_{\tau 3}^2$, $U_{e 1}^2=U_{e
2}^2$, $\Delta m_{12}^2=0.00003eV^2$, and phase $\phi=0$.}
\label{fig_c01}
\end{figure}

\begin{figure}
\centerline{\psfig{file=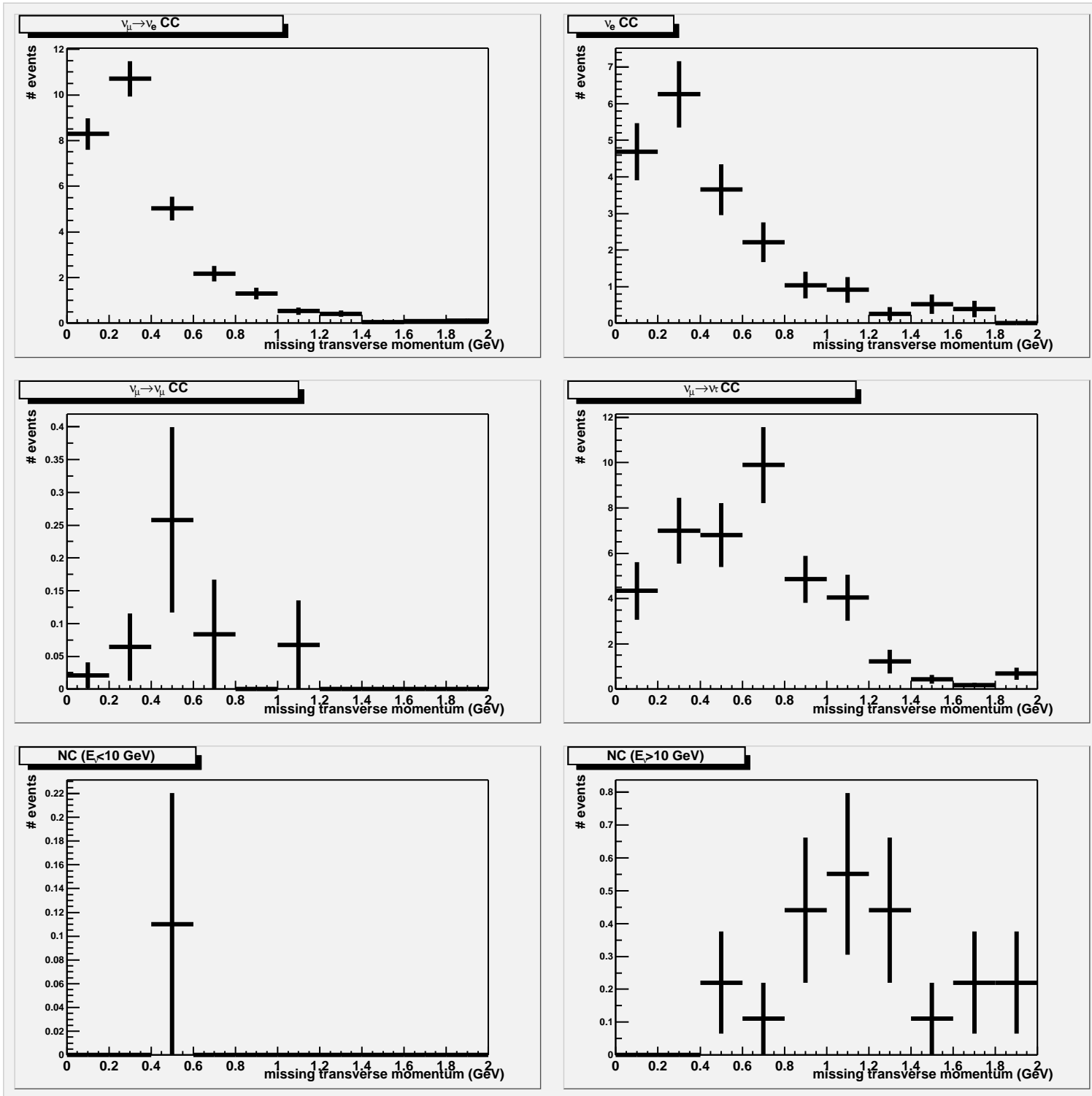}}
\caption{PH2 medium/HED/SLAC event distributions for the missing
transverse momentum ($PT$).  We have used a 20\% track momentum
resolution (from the multiple scattering method) and 1~mrad track
angular resolution.  (Obviously, we have not used neutrinos, neutrons,
or K-longs in the $P_X,P_Y$ sums.)  We have incorporated multiple
coulomb scattering; in order to minimizing the smearing due to
multiple scattering we only use charged particles with momentum
greater than 400~MeV in the sums.  We require missing transverse
momentum $PT < $~400~MeV.  The histograms are normalized to 4x 40
kt-yr data samples.  An exact calculation of the oscillation
probability in matter was used for $\Delta m_{23}^2=0.003eV^2$
(regular mass hierarchy), $|U_{e3}|^2=0.003$ (ie. $\theta_{13}=3^o$),
$U_{\mu 3}^2=U_{\tau 3}^2$, $U_{e 1}^2=U_{e 2}^2$, $\Delta
m_{12}^2=0.00003eV^2$, and phase $\phi=0$.}
\label{fig_c03}
\end{figure}

\begin{figure}
\centerline{\psfig{file=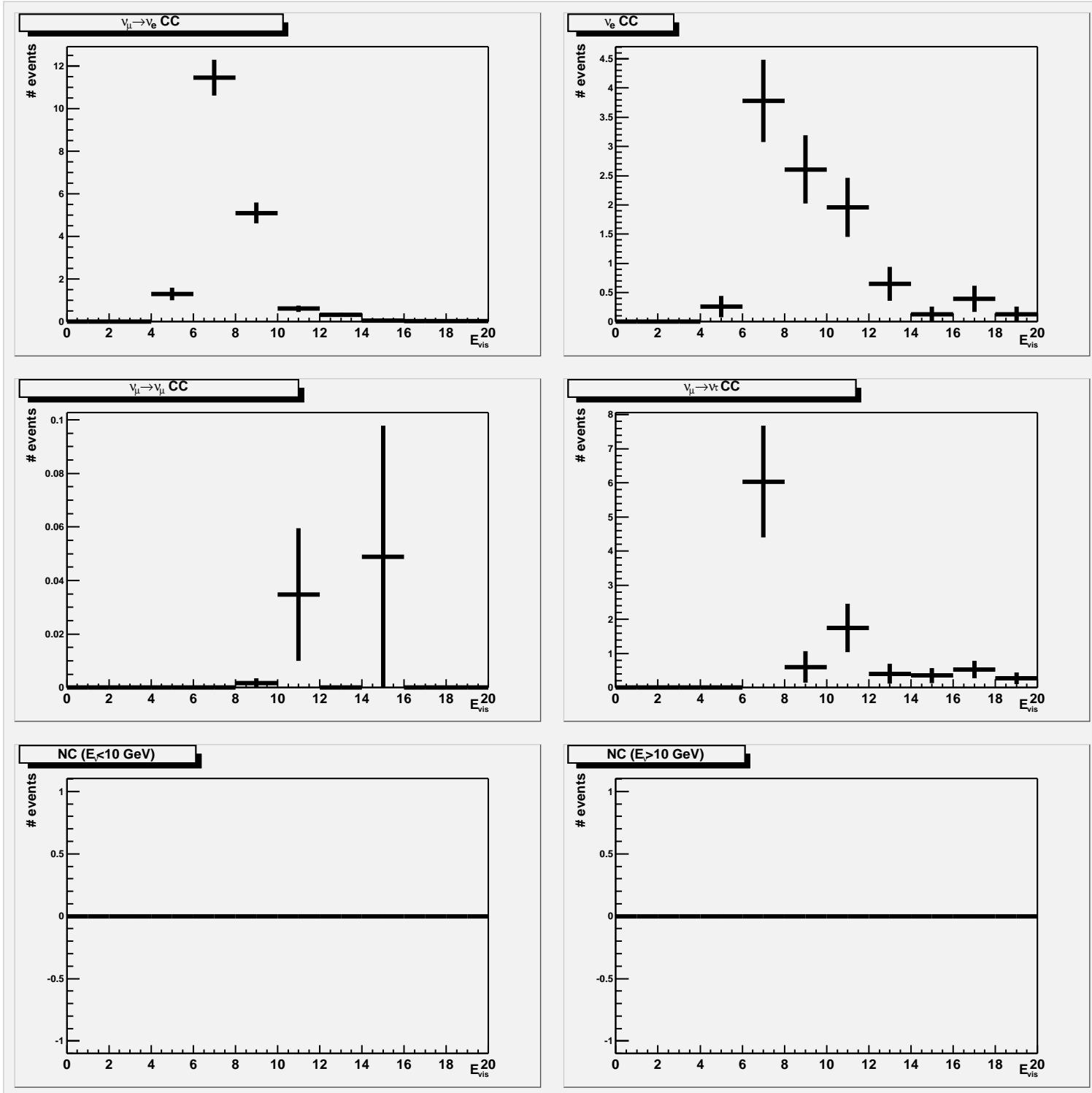}}
\caption{PH2 medium/HED/SLAC event distributions for the visible
energy ($E_{vis}$).  We have used a 20\% track momentum resolution
(from the multiple scattering method) and a 25\%/sqrt(E)
electromagnetic shower energy resolution (from track counting).  We
require $E_{vis} < $ 10 GeV.  The histograms are normalized to 4x 40
kt-yr data samples.  An exact calculation of the oscillation
probability in matter was used for $\Delta m_{23}^2=0.003eV^2$
(regular mass hierarchy), $|U_{e3}|^2=0.003$ (ie. $\theta_{13}=3^o$),
$U_{\mu 3}^2=U_{\tau 3}^2$, $U_{e 1}^2=U_{e 2}^2$, $\Delta
m_{12}^2=0.00003eV^2$, and phase $\phi=0$.}
\label{fig_c05}
\end{figure}

\begin{figure}
\centerline{\psfig{file=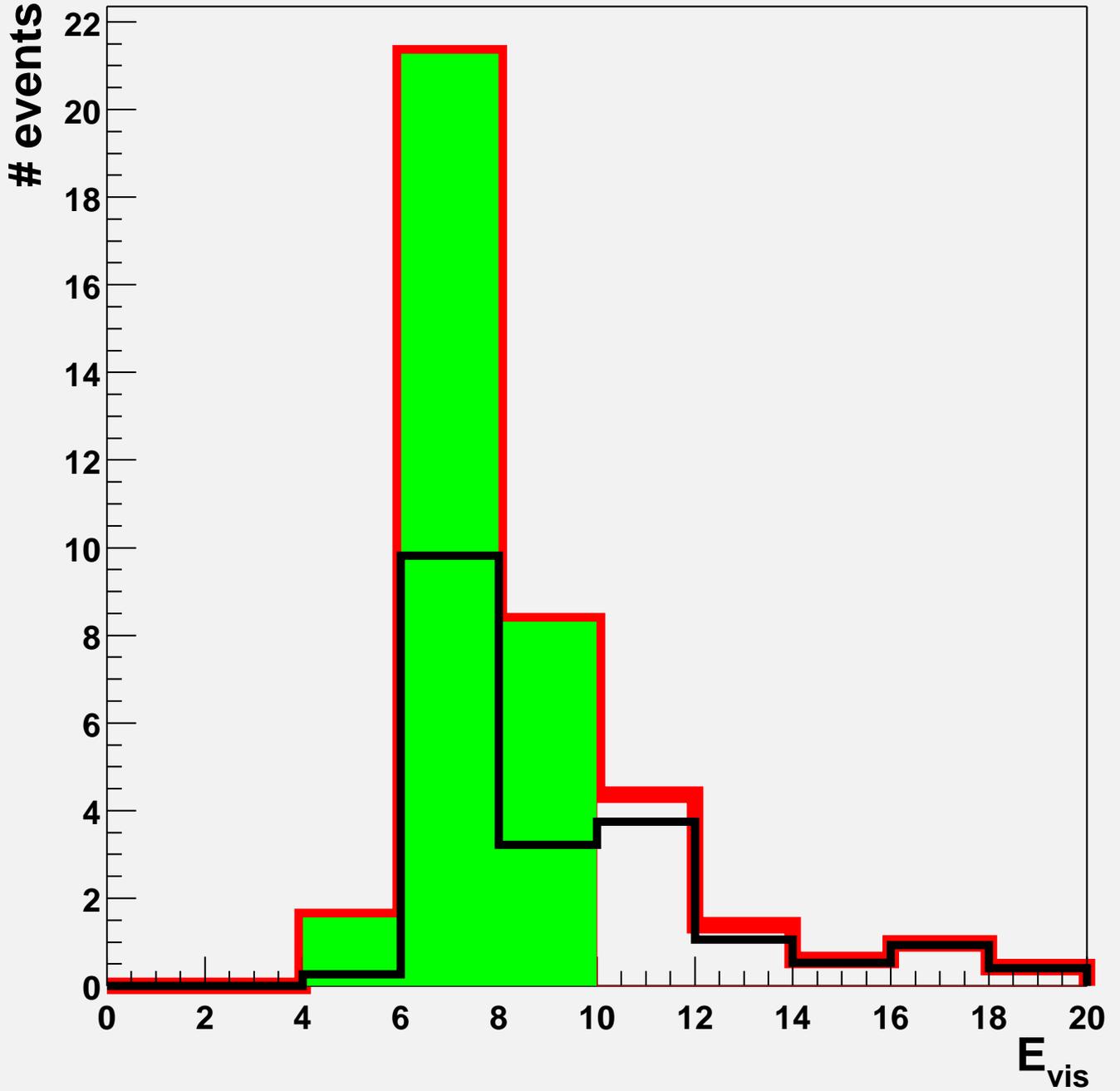}}
\caption{PH2 medium/HED/SLAC total energy event distributions for
background (black line) and background plus signal (red line).  We
require the total energy to be less than 10~GeV (green region).  The
histograms are normalized to 4x 40 kt-yr data samples.  An exact
calculation of the oscillation probability in matter was used for
$\Delta m_{23}^2=0.003eV^2$ (regular mass hierarchy),
$|U_{e3}|^2=0.003$ (ie. $\theta_{13}=3^o$), $U_{\mu 3}^2=U_{\tau
3}^2$, $U_{e 1}^2=U_{e 2}^2$, $\Delta m_{12}^2=0.00003eV^2$, and phase
$\phi=0$.}
\label{fig_c06}
\end{figure}

We use the case of PH2 medium to SLAC as an example.  The data
reduction can be summarized as follows: 1) reject events with
$P_{\mu}>1GeV$, 2) require the highest energy electromagnetic shower
(either from an electron or prompt gamma) to be greater than some
optimized threshold (see figure~\ref{fig_c01}), 3) reject prompt
gammas which do not convert before passing through the first emulsion
sheet after the primary interaction vertex (this rejects 88\% of
prompt gammas), 4) require missing transverse momentum\footnote{Note
that in order to minimize the missing transverse momentum smearing due
to multiple Coulomb scattering we only use charged particles with
momentum greater than 400~MeV in the two independent transverse
momentum sums.} to be less than some optimized threshold (see
figure~\ref{fig_c03}), and 5) require the total visible energy to fall
within an energy range optimized to $\Delta m^2$ (see
figures~\ref{fig_c05} and~\ref{fig_c06}).  The numbers of events after
each sequential cut are tabulated in table~\ref{tab_s02}.

\section{Results}

We assume 40 kt-yr exposures of the detectors in Fermilab main
injector beams whose fluxes have been upgraded by a factor of 4.  We
also assume that the systematic error is dominated by uncertainty in
the number of background events, and that this error is 10\%.  With
these assumptions, we may summarize the significance of the signal in
units of $\sigma_{stat}$ and $\sigma_{syst}$ after all data reduction
cuts in the following table:
\begin{center}
\begin{tabular}{|l||r|r|r|r|r|r|}
\hline 

\multicolumn{7}{|c|}{\bf Significance of the signal after all cuts in
units of $\sigma_{stat}$ ($\sigma_{syst}$)} \\
\multicolumn{7}{|c|}{regular hierarchy, $\Delta m_{23}^2=0.003eV^2$,
$|U_{e3}|^2=0.003$} \\
\hline 
 & \multicolumn{3}{|c|}{\bf MINOS-like detector} &
 \multicolumn{3}{|c|}{\bf OPERA-like detector} \\
 & {\bf Soudan} & {\bf BNL} & {\bf SLAC} & {\bf Soudan} & {\bf BNL} &
 {\bf SLAC} \\
\hline 
\hline 
PH2 low & 1.5 (0.8) & 2.3 (2.1) & 1.4 (2.2) & 2.6 (2.2) & 3.2 (4.3) & 2.0 (5.2) \\ 
\hline 
PH2 medium & 1.1 (0.4) & 1.9 (1.3) & 3.1 (2.6) & 1.4 (1.3) & 2.5 (4.2) & 4.5 (10.0) \\
\hline 
PH2 high & 0.4 (0.1) & 0.7 (0.4) & 1.7 (0.8) & 1.1 (0.8) & 1.9 (2.6) & 3.3 (7.5) \\ 
\hline
\end{tabular}
\end{center}

\begin{figure}
\centerline{\psfig{file=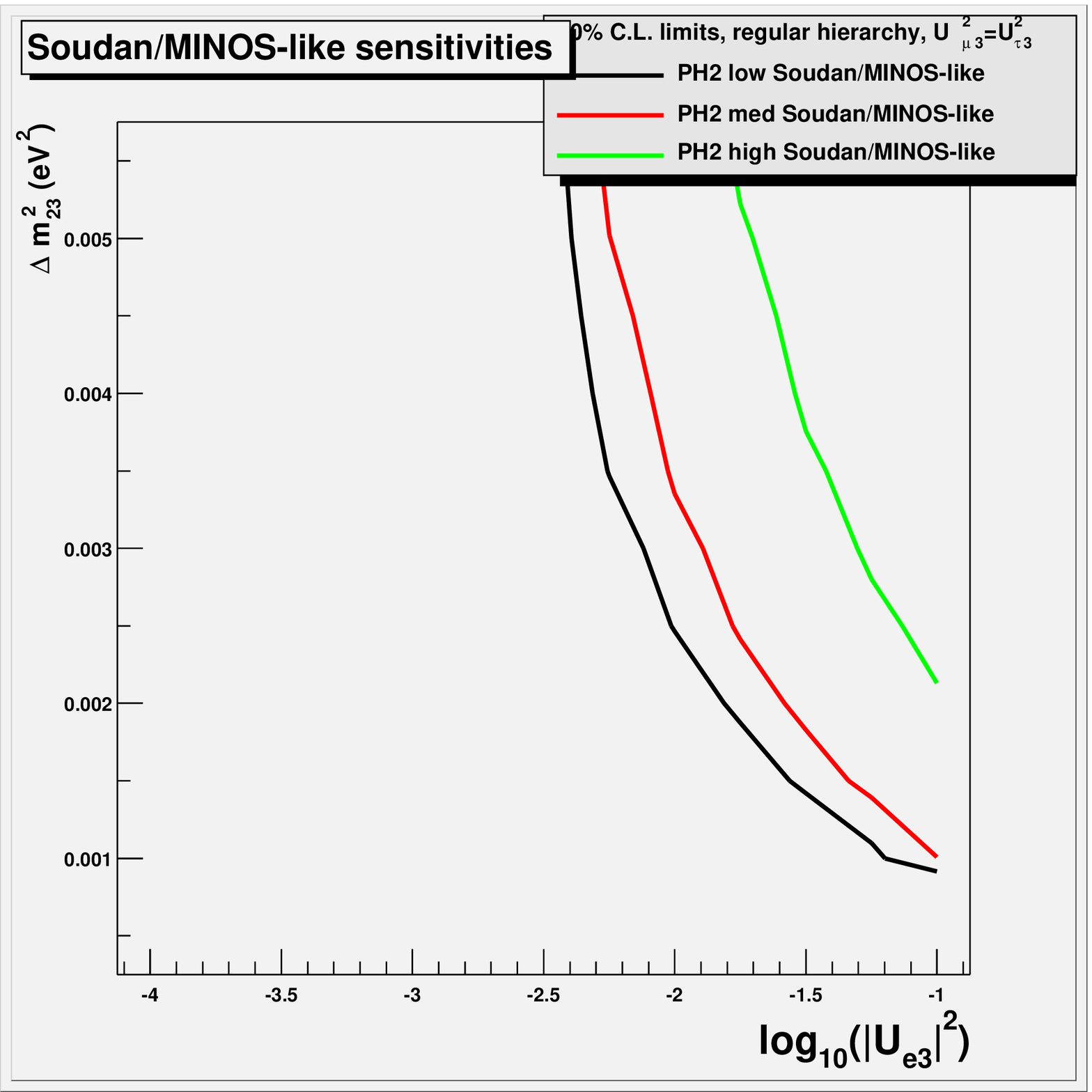}}
\caption{(4x) 40 kt-yr 90\% confidence level limits on $|U_{e3}|^2$
with Fermilab main injector beam pointed at Soudan/MINOS-like.  We have
assumed a systematic error of 10\% on the number of background events.
An exact calculation of the oscillation probability in matter was used
for regular mass hierarchy, $U_{\mu 3}^2=U_{\tau 3}^2$, $U_{e
1}^2=U_{e 2}^2$, $\Delta m_{12}^2=0.00003eV^2$, and phase $\phi=0$.}
\label{fig_b01}
\end{figure}

\begin{figure}
\centerline{\psfig{file=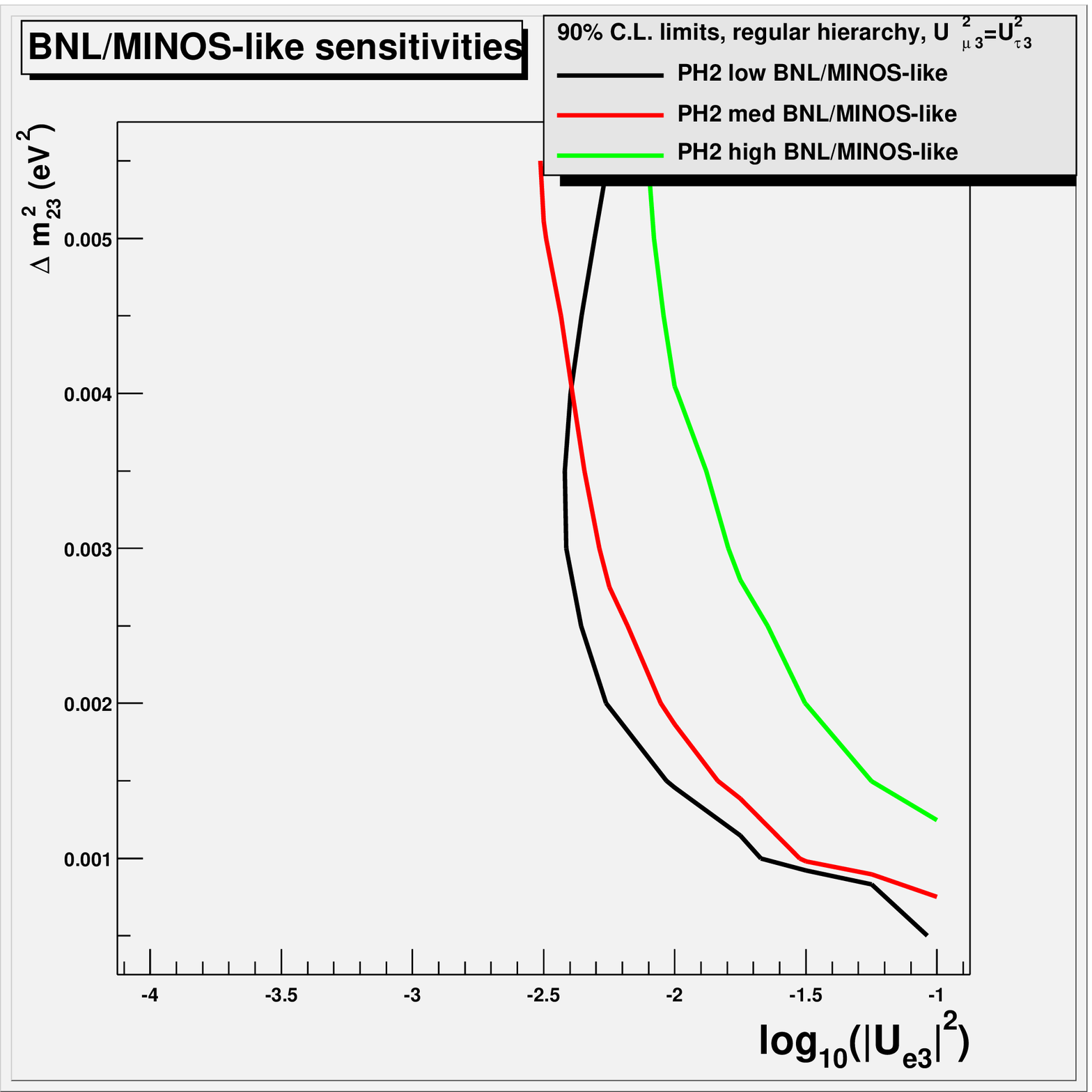}}
\caption{(4x) 40 kt-yr 90\% confidence level limits on $|U_{e3}|^2$
with Fermilab main injector beam pointed at BNL/MINOS-like.  We have
assumed a systematic error of 10\% on the number of background events.
An exact calculation of the oscillation probability in matter was used
for regular mass hierarchy, $U_{\mu 3}^2=U_{\tau 3}^2$, $U_{e
1}^2=U_{e 2}^2$, $\Delta m_{12}^2=0.00003eV^2$, and phase $\phi=0$.}
\label{fig_b02}
\end{figure}

\begin{figure}
\centerline{\psfig{file=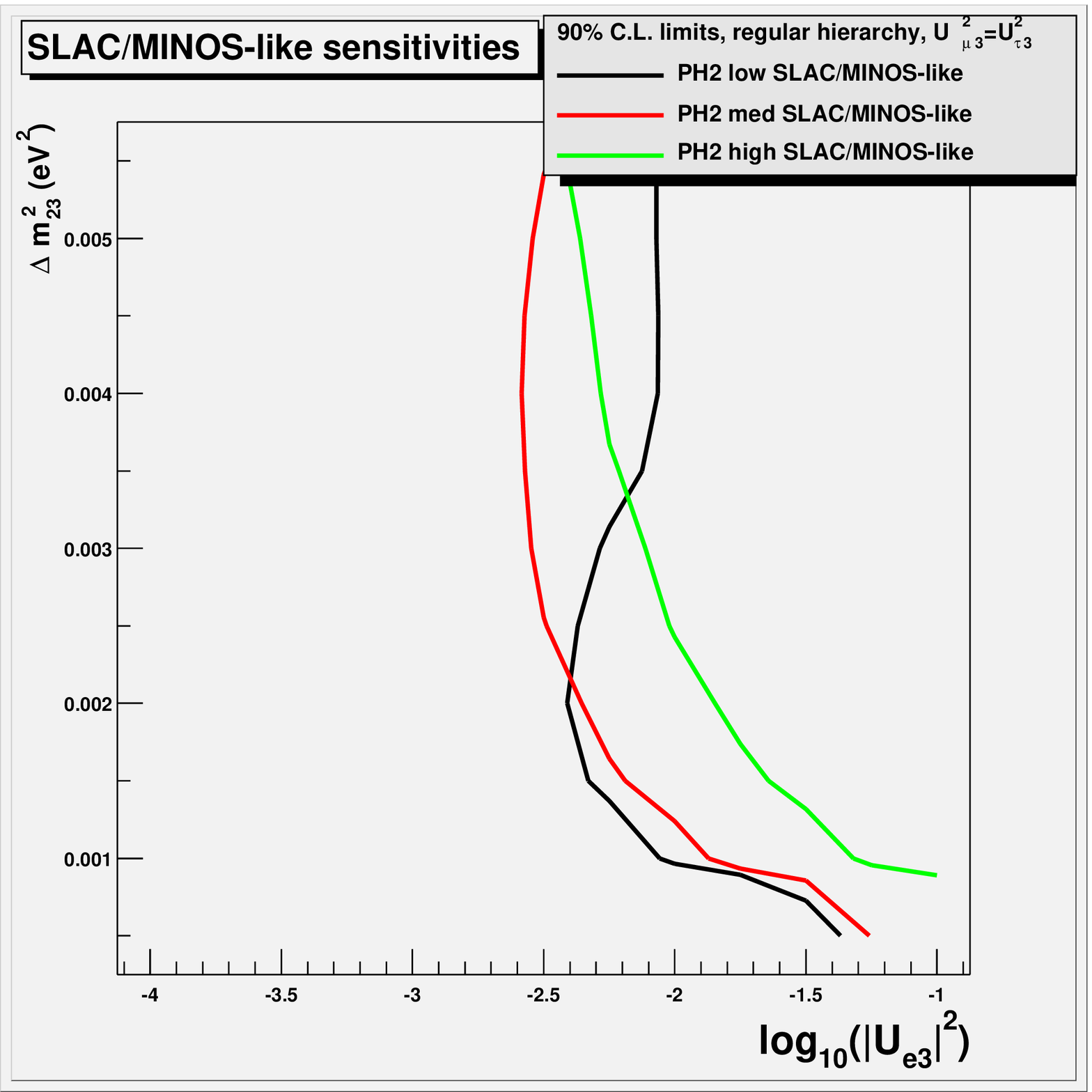}}
\caption{(4x) 40 kt-yr 90\% confidence level limits on $|U_{e3}|^2$
with Fermilab main injector beam pointed at SLAC/MINOS-like.  We have
assumed a systematic error of 10\% on the number of background events.
An exact calculation of the oscillation probability in matter was used
for regular mass hierarchy, $U_{\mu 3}^2=U_{\tau 3}^2$, $U_{e
1}^2=U_{e 2}^2$, $\Delta m_{12}^2=0.00003eV^2$, and phase $\phi=0$.}
\label{fig_b03}
\end{figure}

\begin{figure}
\centerline{\psfig{file=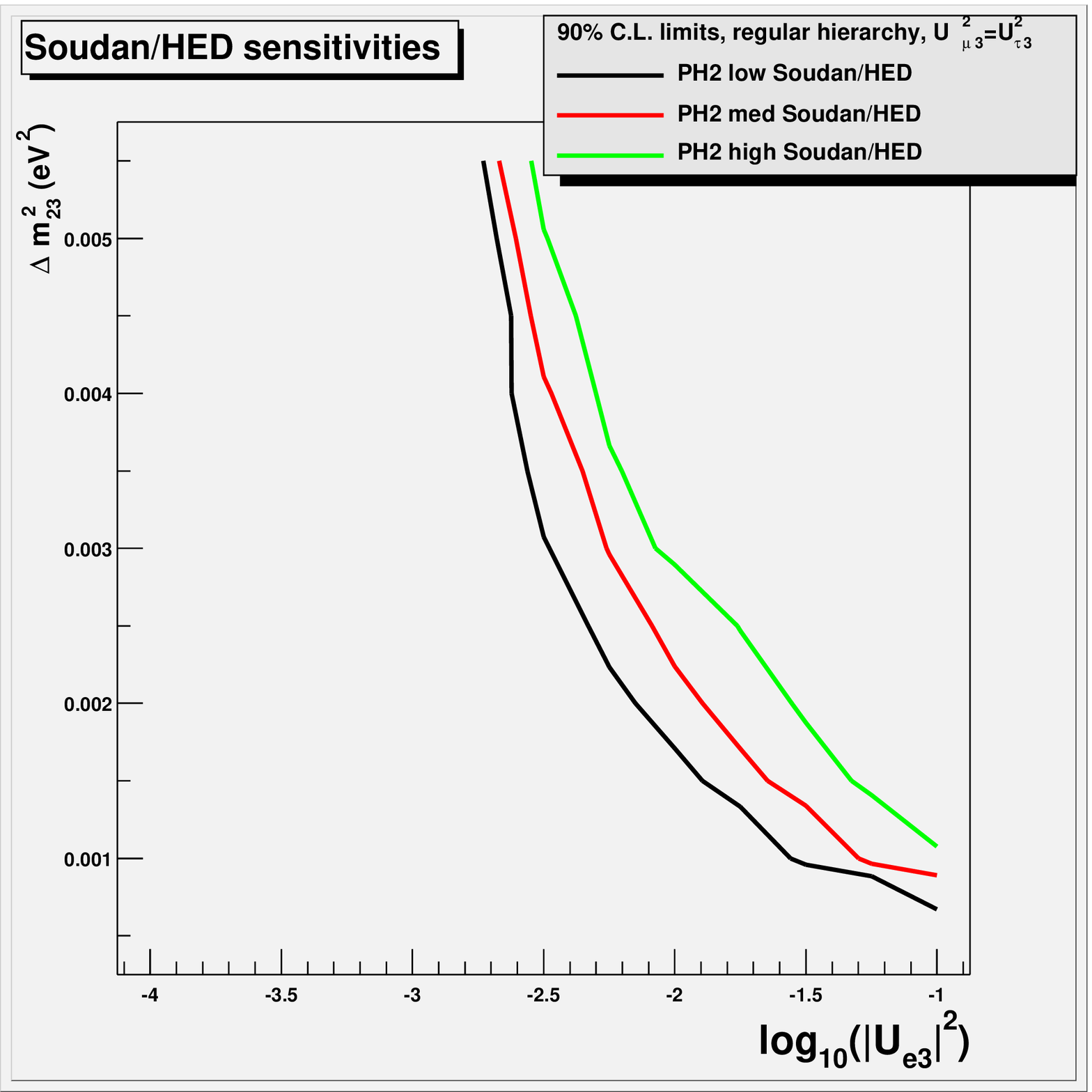}}
\caption{(4x) 40 kt-yr 90\% confidence level limits on $|U_{e3}|^2$
with Fermilab main injector beam pointed at Soudan/HED.  We have
assumed a systematic error of 10\% on the number of background events.
An exact calculation of the oscillation probability in matter was used
for regular mass hierarchy, $U_{\mu 3}^2=U_{\tau 3}^2$, $U_{e
1}^2=U_{e 2}^2$, $\Delta m_{12}^2=0.00003eV^2$, and phase $\phi=0$.}
\label{fig_b04}
\end{figure}

\begin{figure}
\centerline{\psfig{file=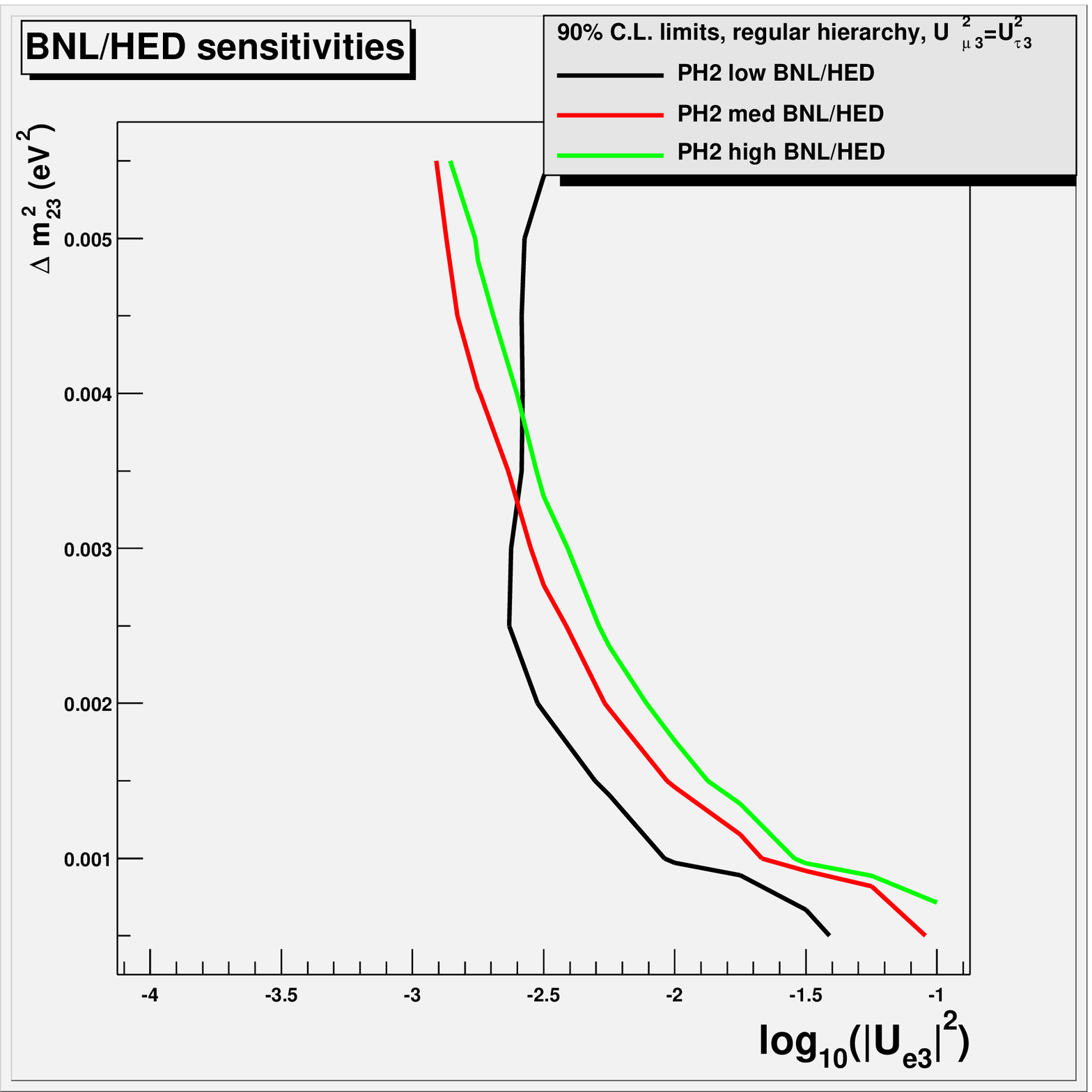}}
\caption{(4x) 40 kt-yr 90\% confidence level limits on $|U_{e3}|^2$
with Fermilab main injector beam pointed at BNL/HED.  We have assumed
a systematic error of 10\% on the number of background events.  An
exact calculation of the oscillation probability in matter was used
for regular mass hierarchy, $U_{\mu 3}^2=U_{\tau 3}^2$, $U_{e
1}^2=U_{e 2}^2$, $\Delta m_{12}^2=0.00003eV^2$, and phase $\phi=0$.}
\label{fig_b05}
\end{figure}

\begin{figure}
\centerline{\psfig{file=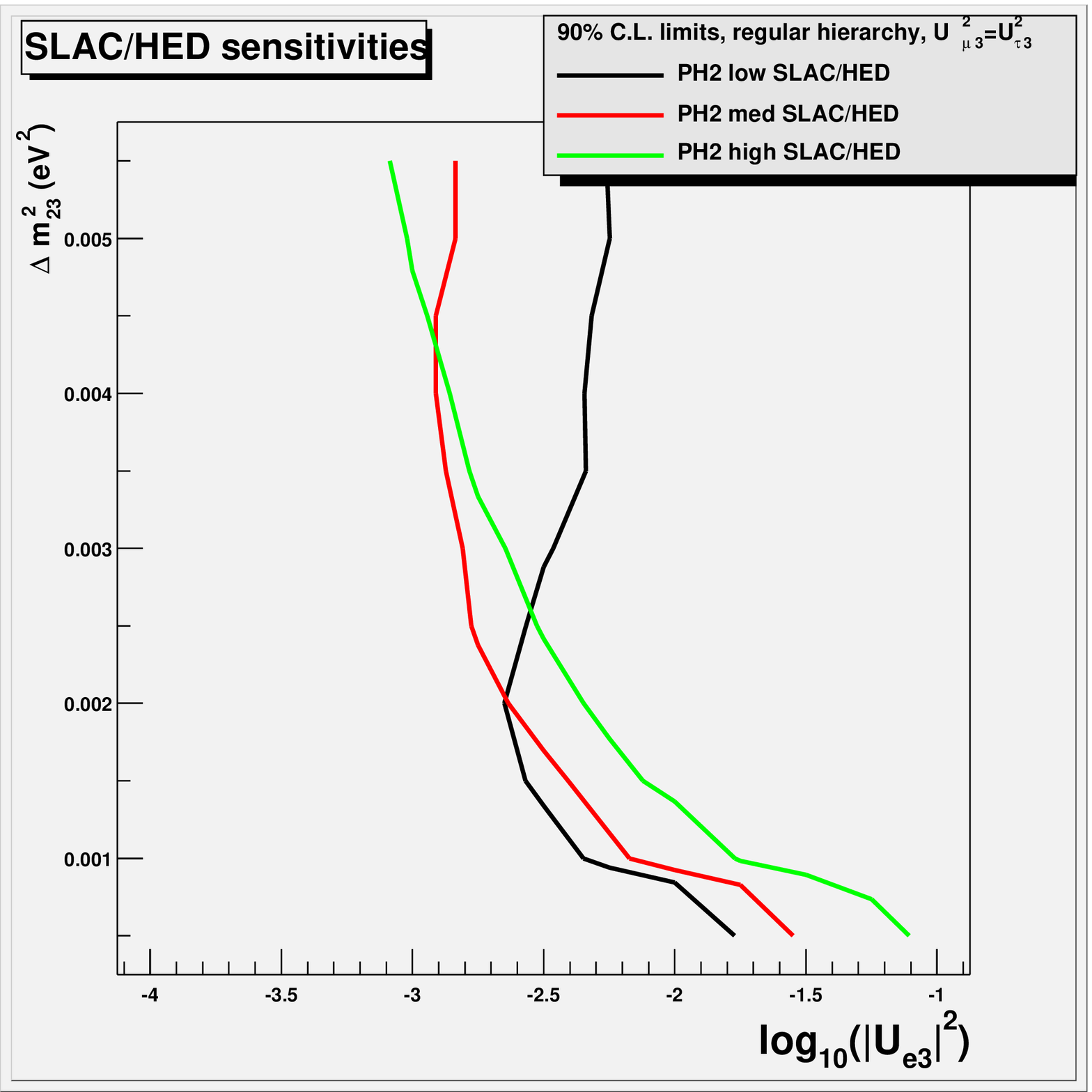}}
\caption{(4x) 40 kt-yr 90\% confidence level limits on $|U_{e3}|^2$
with Fermilab main injector beam pointed at SLAC/HED.  We have assumed
a systematic error of 10\% on the number of background events.  An
exact calculation of the oscillation probability in matter was used
for regular mass hierarchy, $U_{\mu 3}^2=U_{\tau 3}^2$, $U_{e
1}^2=U_{e 2}^2$, $\Delta m_{12}^2=0.00003eV^2$, and phase $\phi=0$.}
\label{fig_b06}
\end{figure}

\begin{figure}
\centerline{\psfig{file=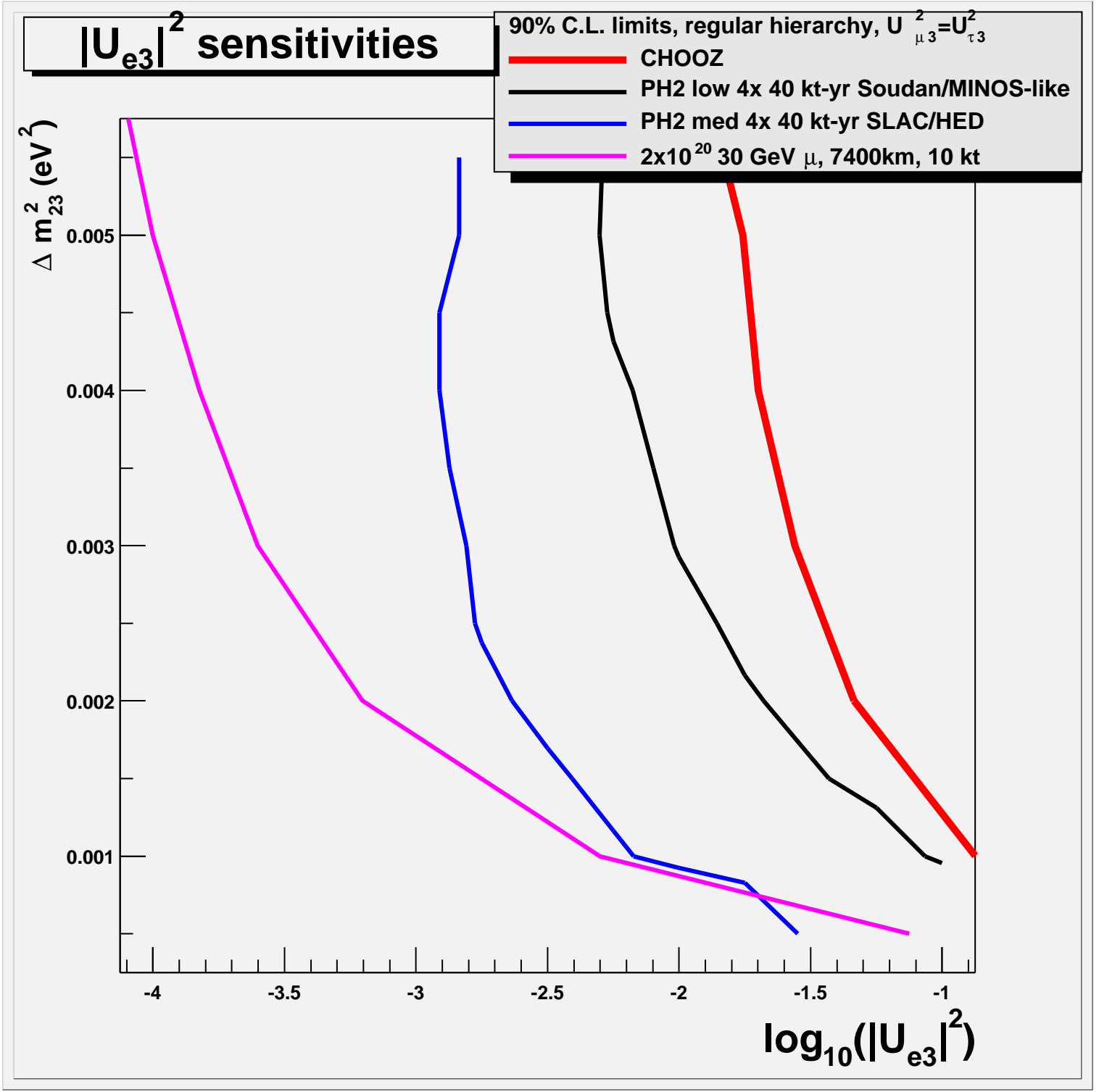}}
\caption{Predicted 90\% confidence level limits on $|U_{e3}|^2$ with
Fermilab main injector beam.  We have assumed a systematic error of
10\% on the number of background events.  An exact calculation of the
oscillation probability in matter was used for regular mass hierarchy,
$U_{\mu 3}^2=U_{\tau 3}^2$, $U_{e 1}^2=U_{e 2}^2$, $\Delta
m_{12}^2=0.00003eV^2$, and phase $\phi=0$.}
\label{fig_b07}
\end{figure}

\begin{figure}
\centerline{\psfig{file=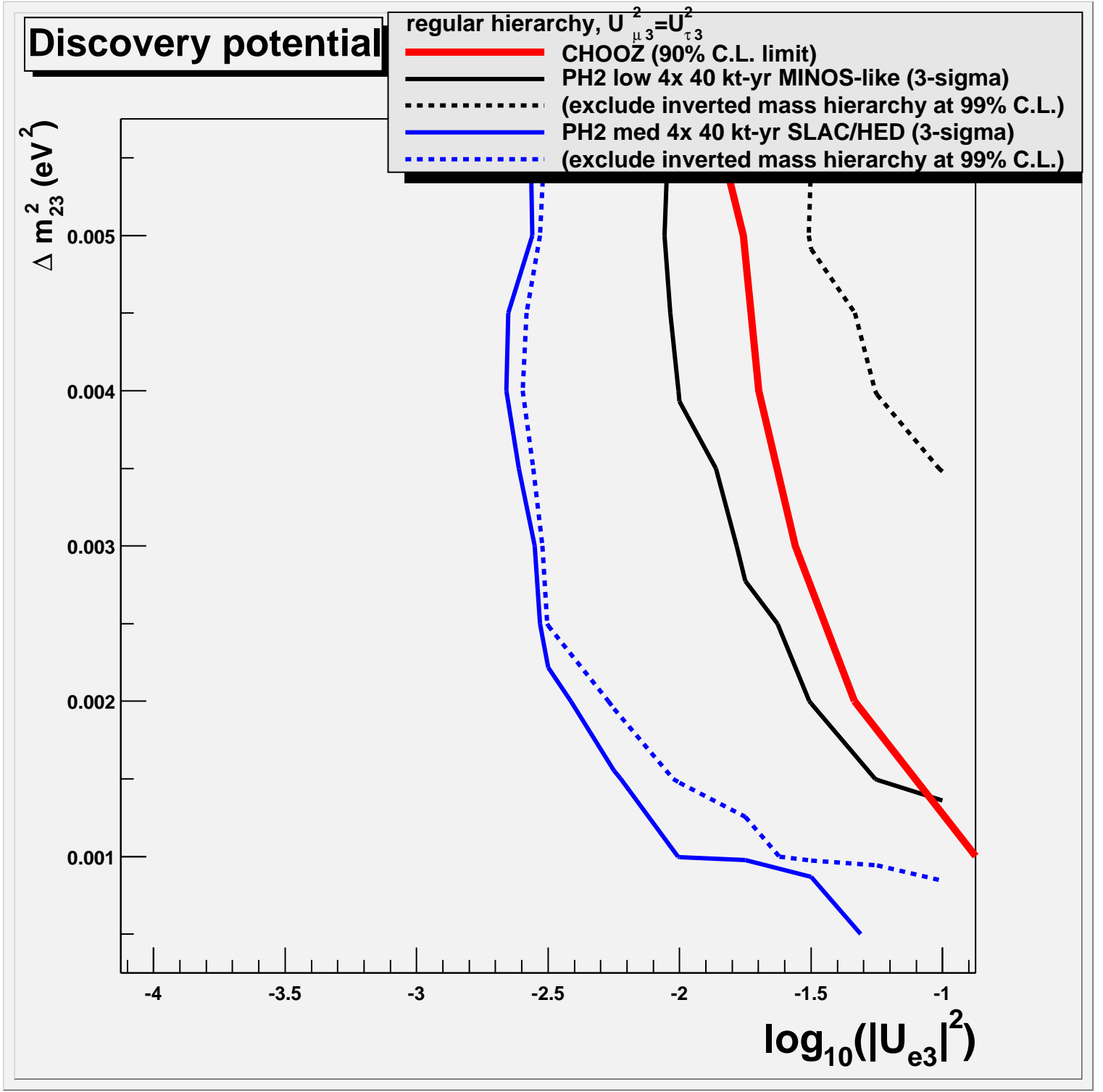}}
\caption{Discovery potential for $|U_{e3}|^2$ and sign($\Delta
m_{23}^2$) with Fermilab main injector beam.  We have assumed a
systematic error of 10\% on the number of background events.  For
sign($\Delta m_{23}^2$) we have assumed an additional sample of data
with the same exposure except with an anti-neutrino beam.  An exact
calculation of the oscillation probability in matter was used for
regular mass hierarchy, $U_{\mu 3}^2=U_{\tau 3}^2$, $U_{e 1}^2=U_{e
2}^2$, $\Delta m_{12}^2=0.00003eV^2$, and phase $\phi=0$.}
\label{fig_b08}
\end{figure}

The predicted sensitivities to $|U_{e3}|^2$ as a function of $\Delta
m_{23}^2$ are shown in figures~\ref{fig_b01} through~\ref{fig_b06}.
Two general conclusions may be made: 1) OPERA-like is better than
MINOS-like, and 2) running ``on the highest energy oscillation peak''
is best.  If we take these calculations at face value, then the best
beam/site/detector setup for $\Delta m_{23}^2=0.003eV^2$ appears to be
the PH2 medium beam pointed at SLAC/HED.  (If $\Delta m_{23}^2$ is
slightly higher or lower, then a slightly higher or lower energy beam
is optimal.)  The predicted sensitivity to $|U_{e3}|^2$ including
statistical and systematic errors for a 40-kt exposure of HED at SLAC
in PH2 medium (4x regular flux) is shown in figure~\ref{fig_b07}.
Similarly, the discovery potential for $|U_{e3}|^2$ and sign($\Delta
m^2_{23}$) is shown in figure~\ref{fig_b08}.  Note that
$\nu_\mu\rightarrow\nu_e$ appearance signals and the sign of $\Delta
m_{23}^2$ can be obtained at the 3-sigma level for
$\theta_{13}\tilde{>}$3~degrees.

We would like to thank Stanley Wojcicki, Vittorio Paolone, and Deborah
Harris for several useful comments and discussions.  Also, we would
like to thank Mark Messier for providing a cross-check of our
implementation of the Barger et.al.~\cite{bar80} method for
calculating neutrino oscillations in matter.

This work was sponsored by the National Science Foundation.

\begin{table}[p]
\begin{center}
\rotatebox{90}{
\begin{tabular}{|l||r||r|r|r|r|r|}
\hline
\multicolumn{7}{|c|}{FNAL-SLAC (2900 km) 40 kt-yr exposure of
MINOS-like detector in PH2 medium (4x)} \\
\hline
 & \multicolumn{1}{c||}{signal} & \multicolumn{5}{c|}{background} \\
\hline
 &\multicolumn{1}{c||}{CC}&\multicolumn{3}{c|}{CC}&\multicolumn{2}{c|}{NC}\\
sequential cuts:&$\nu_\mu\rightarrow\nu_e$&beam $\nu_e$&$\nu_\mu\rightarrow\nu_\mu$&$\nu_\mu\rightarrow\nu_\tau$&$E_\nu<10GeV$&$E_\nu>10GeV$\\
\hline 
\hline
all events & 126.4 & 130.4 & 4589.8 & 1325.7 & 3312.7 & 1279.3 \\
\hline
$P_\mu<1GeV$ & 126.4 & 130.4 & 374.9 & 1127.6 & 3312.7 & 1279.3 \\
\hline
$E_{CLUST}/E_{TOT}>0.7$ & 59.3 & 68.2 & 18.8 & 278.2 & 753.1 & 203.0 \\
\hline
cluster $N_{STRIPS}\geq 9$ & 57.9 & 67.4 & 16.5 & 230.2 & 135.9 & 94.6 \\
\hline
neural net $Y>0$ & 52.0 & 60.7 & 8.8 & 173.6 & 86.3 & 71.6 \\
\hline
$400pe<E_{TOT}<800pe$ & 34.9 & 12.3 & 3.0 & 82.8 & 29.4 & 15.6 \\
\hline
\hline
efficiency, bkgd fraction& 0.28 & 0.094 & 0.0007 & 0.062 & 0.009 & 0.012 \\
\hline
\end{tabular}}
\caption{Summary of data reduction for 40 kt-yr exposure of a MINOS-like
detector at SLAC in the PH2 medium beam (4x).  An exact calculation of the
oscillation probability in matter was used for $\Delta
m_{23}^2=0.003eV^2$ (regular hierarchy), $|U_{e3}|^2=0.003$
(ie. $\theta_{13}=3^o$), $U_{\mu 3}^2=U_{\tau 3}^2$, $\Delta
m_{12}^2=0.00003eV^2$, $U_{e1}^2=U_{e2}^2$, and phase $\phi=0$.}
\label{tab_s01}
\end{center}
\end{table}

\begin{table}[p]
\begin{center}
\rotatebox{90}{
\begin{tabular}{|l||r||r|r|r|r|r|}
\hline
\multicolumn{7}{|c|}{FNAL-SLAC (2900 km) 40 kt-yr exposure of HED detector in PH2 medium (4x)} \\
\hline
 & \multicolumn{1}{c||}{signal} & \multicolumn{5}{c|}{background} \\
\hline
 &\multicolumn{1}{c||}{CC}&\multicolumn{3}{c|}{CC}&\multicolumn{2}{c|}{NC}\\
sequential cuts:&$\nu_\mu\rightarrow\nu_e$&beam $\nu_e$&$\nu_\mu\rightarrow\nu_\mu$&$\nu_\mu\rightarrow\nu_\tau$&$E_\nu<10GeV$&$E_\nu>10GeV$\\
\hline 
\hline
all events & 126.4 & 130.4 & 4589.8 & 1325.7 & 3312.7 & 1279.3 \\
\hline
$P_\mu<1GeV$ & 126.4 & 130.4 & 374.9 & 1127.6 & 3312.7 & 1279.3 \\
\hline
$E_{e,\gamma}>5GeV$ & 28.9 & 20.6 & 4.8 & 39.7 & 0.9 & 28.5 \\
\hline
$\gamma$ conversion & 28.9 & 20.6 & 0.6 & 39.7 & 0.1 & 3.4 \\
\hline
missing $PT<400MeV$ & 19.0 & 11.0 & 0.09 & 11.3 & 0 & 0 \\
\hline
$E_{vis}<10GeV$ & 17.9 & 6.6 & 0.002 & 6.6 & 0 & 0 \\
\hline
\hline
efficiency, bkgd fraction& 0.14 & 0.051 & 0 & 0.005 & 0 & 0 \\
\hline
\end{tabular}}
\caption{Summary of data reduction for 40 kt-yr exposure of a
HED detector at SLAC in the PH2 medium beam (4x).  An exact
calculation of the oscillation probability in matter was used for
$\Delta m_{23}^2=0.003eV^2$ (regular hierarchy), $|U_{e3}|^2=0.003$
(ie. $\theta_{13}=3^o$), $U_{\mu 3}^2=U_{\tau 3}^2$, $\Delta
m_{12}^2=0.00003eV^2$, $U_{e1}^2=U_{e2}^2$, and phase $\phi=0$.}
\label{tab_s02}
\end{center}
\end{table}

\newpage

\end{document}